\documentclass[aps,pre,reprint,groupedaddress,longbibliography,nobalancelastpage]{revtex4-2}
\usepackage{amsmath,amssymb,amsthm,dcolumn}
\usepackage{graphicx}
\usepackage{xcolor}
\usepackage{lmodern}
\usepackage{orcidlink}

\usepackage{hyperref}
\usepackage{stmaryrd}
\usepackage{enumitem}
\usepackage{accents}
\usepackage{CJK}
\renewcommand{\vec}[1]{\boldsymbol{#1}}

\usepackage[compact]{titlesec}

\DeclareMathAlphabet{\mathbfsfit}{\encodingdefault}{\sfdefault}{bx}{sl}
\newcommand{\tens}[1]{\mathbfsfit{#1}}

\renewcommand{\eqref}[1]{\hyperref[#1]{(\ref*{#1})}}
\newcommand{\figref}[2]{[Fig.~\hyperref[#1]{\ref*{#1}(#2)}]}
\newcommand{\bfigref}[3]{[Fig.~\hyperref[#1]{\ref*{#1}(#2\textsubscript{#3})}]}
\newcommand{\figrefi}[2]{[Fig.~\hyperref[#1]{\ref*{#1}(#2)}, inset]}
\newcommand{\textfigref}[2]{Fig.~\hyperref[#1]{\ref*{#1}(#2)}}
\newcommand{\textfigureref}[2]{Figure~\hyperref[#1]{\ref*{#1}(#2)}}
\newcommand{\wholefigref}[1]{(Fig.~\ref{#1})}

\newcommand{\figrefp}[2]{\hyperref[#1]{\ref*{#1}(#2)}}

\DeclareMathAlphabet{\mathcalbf}{OMS}{cmsy}{b}{n}
\DeclareMathAlphabet{\mathbfsf}{\encodingdefault}{\sfdefault}{b}{n}
\DeclareMathOperator{\tr}{tr}

\DeclareFontEncoding{LGR}{}{}
\DeclareSymbolFont{sfgreek2}{LGR}{cmss}{bx}{it}
\DeclareMathSymbol{\teps}{\mathord}{sfgreek2}{`e}

\definecolor{linkcolor}{HTML}{223096}
\hypersetup{colorlinks,allcolors=linkcolor}

\bibpunct{\textcolor{linkcolor}{[}}{\textcolor{linkcolor}{]}}{\textcolor{linkcolor}{,}}{n}{}{;}
\renewcommand{\i}{\mathrm{i}}
\newcommand{\e}{\mathrm{e}}
\newcommand{\Psii}{\mathit{\Psi}}

\begin{document}
\begin{CJK*}{UTF8}{gbsn}

\title{Stretching and bending of (really) thick elastic plates}

\author{Shiheng Zhao (赵世恒)\,\orcidlink{0009-0003-0024-5999}}
\affiliation{Max Planck Institute for the Physics of Complex Systems, N\"othnitzer Stra\ss e 38, 01187 Dresden, Germany}
\affiliation{\mbox{Max Planck Institute of Molecular Cell Biology and Genetics, Pfotenhauerstra\ss e 108, 01307 Dresden, Germany}}
\affiliation{Center for Systems Biology Dresden, Pfotenhauerstra\ss e 108, 01307 Dresden, Germany}
\author{Pierre A. Haas\,\orcidlink{0000-0002-6663-0393}}
\email[Contact author: ]{haas@pks.mpg.de}
\affiliation{Max Planck Institute for the Physics of Complex Systems, N\"othnitzer Stra\ss e 38, 01187 Dresden, Germany}
\affiliation{\mbox{Max Planck Institute of Molecular Cell Biology and Genetics, Pfotenhauerstra\ss e 108, 01307 Dresden, Germany}}
\affiliation{Center for Systems Biology Dresden, Pfotenhauerstra\ss e 108, 01307 Dresden, Germany}
\date{\today}

\begin{abstract}
The mechanical energy of an elastic plate separates into stretching and bending energies. This is a result for asymptotically thin plates, but it is often a surprisingly accurate approximation for thick plates, too. Here, we address this conundrum: We compute the deformations of a thick elastic plate resulting from imposed, asymptotically small deformations of its midline to discover effective stretching and bending moduli. They soften with increasing plate thickness, but, strikingly, their ratio remains approximately constant. In this way, our calculations provide a justification for applying the thin-plate picture of stretching and bending to thick plates such as biological cell sheets.
\end{abstract}
\maketitle
\end{CJK*}

\begin{figure}[b]
\centering
\includegraphics[width=\linewidth]{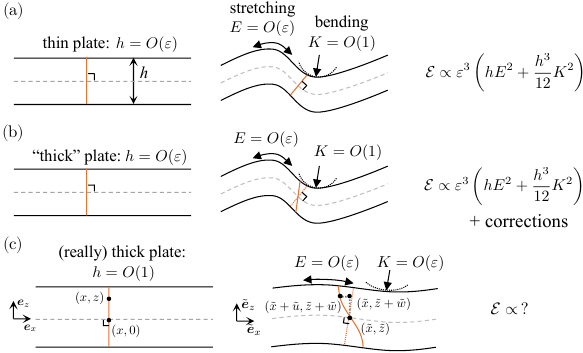}
\caption{Plate theories. (a)~For a thin plate of thickness ${h=O(\varepsilon)}$, where $\varepsilon\ll 1$, the assumptions of small stretching $E=O(\varepsilon)$ and finite bending $K=O(1)$ of the midsurface imply that the elastic energy $\mathcal{E}$ separates into stretching and bending energies at leading order. (b)~Classical theories of ``thick'' plates: including, e.g., shearing of normals introduces corrections to this leading-order energy for the same scaling assumptions. (c)~Small deformations of a (really) thick plate, $h=O(1)$. Points at position $(x,z)$ in the undeformed configuration (left) are displaced by distances $\tilde{u}(x, z)$ and $\tilde{w}(x,z)$ parallel and perpendicular to the undeformed midsurface, respectively, in the deformed configuration (right).}
\label{fig1}
\end{figure}

Slender elastic objects, such as rods, plates, or shells, are described by effective theories that reduce their three-dimensional mechanics to one-dimensional midlines or two-dimensional midsurfaces \mbox{\cite{Love1944,landaulifshitz,audoly08}}. In particular, the elastic energy $\mathcal{E}$ of a thin plate of thickness $h$ separates into stretching and bending energies,
\begin{align}
\mathcal{E}\propto h\left(E^2+\dfrac{h^2}{12}K^2\right),\label{eq:plateE}
\end{align}
where $E$ and $K$ measure the stretching and bending of the midsurface, respectively, and the proportionality hides the material parameters. This is the leading-order term~\cite{Ciarlet1980,audoly08,Steigmann2013,Wang2016} in an expansion, in the small parameter $\varepsilon\ll 1$, of the full, three-dimensional elastic energy of the plate for $h=O(\varepsilon)$, with the midsurface scaling assumptions $E=O(\varepsilon)$, $K=O(1)$ \figref{fig1}{a}. 

This asymptotic understanding emerged from a tradition of phenomenological plate theories associated with the likes of Kirchhoff and Love~\cite{love1888}, Reissner~\cite{Reissner1945}, Mindlin~\cite{Mindlin1951}, and Koiter~\cite{Koiter1966}. For example, the Kirchhoff ``hypothesis'', that normals to the undeformed midsurface remain normals~\figref{fig1}{a}, is an asymptotic consequence of these scaling assumptions. Their formal derivation and proofs of convergence remain an active area of research in applied mathematics~\cite{Fox1993,LeDret1996,Friesecke2006,Neukamm2013}.

In parallel, descriptions of ``thick'' plates have emerged: for example, Reissner--Mindlin plate theory~\cite{Reissner1945,Mindlin1951,Reddy2006} and related theories~\cite{Hencky1947,Kromm1955,Wang1997} break the Kirchhoff hypothesis by allowing shear deformations of the normals~\figref{fig1}{b}. In the context of elastic rods, these shear deformations are included in Cosserat theories~\cite{Antman1995}. Asymptotically, such theories are still descriptions of thin plates \footnote{The equations of thick plates can also be solved in terms of infinite series: see, e.g., Refs.~\cite{Piltner1988,Piltner1992}. We do not refer to these results as ``plate theories'' because they do not the reduce the mechanics to an effective midplane model.}, including corrections to the leading-order energy~\eqref{eq:plateE}.\nocite{Piltner1988,Piltner1992}

Nonetheless, this picture of stretching and bending often provides a very accurate approximation of the mechanics of thick plates far beyond the asymptotic regime in which this picture emerges. For example, motivated by the mechanics of cysts of cells, we have recently confirmed this for the classical Pogorelov dimple problem~\cite{pogorelov,landaulifshitz} on a thick shell~\cite{zhao25}.

Here, we therefore ask: does the separation into stretching and bending energies also emerge for really thick plates? We address this question by solving the equations of elasticity for small deformations, $E=O(\varepsilon)$, $K=O(\varepsilon)$, of a plate with $h=O(1)$ \figref{fig1}{c}. We compute effective stretching and bending moduli of this plate and suggest an explanation for the unearned accuracy of thin-plate theories by showing that their ratio is approximately independent of plate thickness.

\vspace{-5pt}
\paragraph*{Plate kinematics.} We start by describing the kinematics of the thick plate. We describe its undeformed configuration using Cartesian coordinates $x$ along and ${z\in[-h/2,h/2]}$ perpendicular to the midsurface ${z=0}$ \figref{fig1}{c}. The point $(x,0)$ of the undeformed midsurface maps to the point $\bigl(\tilde{x}(x),\tilde{z}(x)\bigr)$ in the deformed configuration. The image of the point $(x,z)$ in the deformed configuration is then $\vec{\tilde{r}}(x,z)=\bigl(\tilde{x}(x)+\tilde{u}(x,z),\tilde{z}(x)+\tilde{w}(x,z)\bigr)$, where $\tilde{u}(x,z)$ and $\tilde{w}(x,z)$ are parallel and perpendicular displacements~\figref{fig1}{c}, with $\tilde{u}(x,0)=\tilde{w}(x,0)=0$.

We introduce the stretch $\tilde{f}(x)$ of and the tangent angle $\tilde{\psi}(x)$ to the deformed midsurface, which satisfy~\cite{Haas21}
\begin{align}
&\tilde x'(x)=\tilde{f}(x)\cos{\tilde{\psi}(x)},&&\tilde z'(x)=\tilde{f}(x)\sin{\tilde{\psi}(x)},
\end{align}
where dashes denote differentiation with respect to the argument, $x$. With this, the curvature of the deformed midsurface is $\tilde{\kappa}(x)=\tilde{\psi}'(x)/\tilde{f}(x)$~\cite{Haas21}.

Let $\{\vec{e_x},\vec{e_z}\}$ be the Cartesian axes of the undeformed configuration. The deformation gradient tensor is defined to be $\tens{F}={\partial \vec{\tilde{r}}}/{\partial x}\otimes \vec{e_x} + {\partial \vec{\tilde{r}}}/{\partial z}\otimes \vec{e_z}$~\cite{goriely}. Explicitly,
\begin{align}
\tens{F}
&=\left(\tilde f\cos{\tilde{\psi}}+\frac{\partial\tilde{u}}{\partial x}\right)\vec{\tilde{e}_x}\otimes \vec{e_x}
+\frac{\partial \tilde{u}}{\partial z}\vec{\tilde{e}_x}\otimes \vec{e_z} \nonumber\\
&\quad+\left(\tilde f\sin{\tilde{\psi}}+\frac{\partial\tilde{w}}{\partial x}\right)\vec{\tilde{e}_z}\otimes \vec{e_x}
+\frac{\partial\tilde{w}}{\partial z}\vec{\tilde{e}_z}\otimes \vec{e_z},\label{eq:F}
\end{align}
with $\{\vec{\tilde e_x},\vec{\tilde e_z}\}$ the Cartesian axes of the deformed configuration, highlighting that $\tens{F}$ is a two-point tensor~\cite{ogden}.

\paragraph*{Plate mechanics.} We assume the simplest, incompressible, \emph{two-dimensional} neo-Hookean constitutive relations~\cite{goriely,* [{For the two-dimensional neo-Hookean constitutive relations, see, e.g., }] [{, }] Zhao2021,*Zhao2026} for the plate. Its elastic energy is therefore
\begin{align}
\mathcal{E}=\iint_{\mathcal{A}}{e\,\mathrm{d}x\,\mathrm{d}z},\quad\text{where }e=\dfrac{C}{2}\bigl(\tr{\tens{F}^ \top\tens{F}}-2\bigr),
\label{eq:E}
\end{align}
in which $C>0$ is a material parameter. We will nondimensionalise $C=1$ hereinafter. The Cauchy stress is~\cite{goriely}
\begin{align}
\tens{T}=\tens{FF}^\top-p\tens{I},
\end{align}
where $\tens{I}$ is the identity, and the pressure $p$ is the Lagrange multiplier enforcing incompressibility, $\det{\tens{F}}=1$. The corresponding first Piola--Kirchhoff stress is $\tens{P}=\tens{T}\tens{F}^{-\top}$, with respect to which the equation of mechanical equilibrium is~\cite{goriely}
\begin{align}
\operatorname{Div}\tens{P}^{\top}=\vec{0},\label{eq:Cauchy}
\end{align}
in which $\operatorname{Div}$ is the divergence operator with respect to the undeformed configuration. In this configuration, the surfaces of the plate are $z=\pm h/2$, with outward normals $\vec{n}^{\pm}=\pm \vec{e_z}$. The force-free boundary conditions on these surfaces of the deformed plate are therefore~\cite{goriely,Haas21}
\begin{align}
\tens{P}(x,\pm h/2)\cdot\vec{e_z}=\vec{0}.\label{eq:CauchyBC}
\end{align}

\paragraph*{Asymptotic expansion.} We solve the equations of mechanical equilibrium for imposed, small deformations of the midline, which we define by letting
\begin{align}
&\tilde{f}(x)=1+\varepsilon E(x), &&\tilde{\psi}(x) = \varepsilon \Psii(x),
\end{align}
for $\varepsilon\ll 1$. We introduce regular expansions of the displacements, viz.,
\begin{subequations}
\begin{align}
\tilde{u}(x,z) &= \varepsilon U_{(1)}(x,z) + \varepsilon^2U_{(2)}(x,z) + O\bigl(\varepsilon^3\bigr), \\
\tilde{w}(x,z) &= z + \varepsilon W_{(1)}(x,z)+\varepsilon^2W_{(2)}(x,z)+O\bigl(\varepsilon^3\bigr).
\end{align}
\end{subequations}
We introduce these expansions into Eq.~\eqref{eq:F}. Using \textsc{Mathematica} (Wolfram, Inc.), we expand the incompressibility condition $\det{\tens{F}}=1$ to obtain~\footnote{See Supplemental Material at [url to be inserted] for a \textsc{Mathematica} notebook implementing the asymptotic calculations.}
\begin{subequations}\label{eq:ic}
\begin{align}
&\frac{\partial U_{(1)}}{\partial x}+\frac{\partial W_{(1)}}{\partial z}+E= 0,\label{eq:ic1}\\
&\frac{\partial U_{(2)}}{\partial x}+\frac{\partial W_{(2)}}{\partial z}-\frac{\partial U_{(1)}}{\partial z}\left(\Psii+\frac{\partial W_{(1)}}{\partial x}\right)\nonumber\\
&\qquad+\frac{\partial W_{(1)}}{\partial z}\left(E+\frac{\partial U_{(1)}}{\partial x}\right)-\dfrac{\Psii^2}{2}= 0.
\end{align}
\end{subequations}
Using these expansions, we find
\begin{align}\label{eq:energy}
e&= \frac{\varepsilon^2}{2}\left[4\left(\frac{\partial U_{(1)}}{\partial x}+E\right)^2+\left(\Psii+\frac{\partial U_{(1)}}{\partial z}+\frac{\partial W_{(1)}}{\partial x}\right)^2\right]\nonumber\\
&\qquad+O\bigl(\varepsilon^3\bigr).
\end{align}
Importantly, this result is independent of $U_{(2)},W_{(2)}$: we have expressed the energy density to leading (second) order in terms of the solution of the first-order mechanical problem only. We solve this problem by expanding
\begin{align}
p(x,z)&=p_{(0)}(x,z) + \varepsilon p_{(1)}(x,z)+O\bigl(\varepsilon^2\bigr).
\end{align}
At leading order, Eqs.~\eqref{eq:Cauchy} and~\eqref{eq:CauchyBC} give
\begin{align}
&\dfrac{\partial p_{(0)}}{\partial x} = \dfrac{\partial p_{(0)}}{\partial z} = 0\quad\text{subject to }p_{(0)}(x,\pm h/2)=1.
\end{align}
Hence $p_{(0)}(x,z)=1$. With this, using Eq.~\eqref{eq:ic1}, the Cauchy equation at next order simplifies to
\begin{subequations}\label{eq:equil}
\begin{align}
    \dfrac{\partial^2U_{(1)}}{\partial x^2} +\dfrac{\partial^2U_{(1)}}{\partial z^2} + \dfrac{\mathrm{d} E}{\mathrm{d} x}-\dfrac{\partial p_{(1)}}{\partial x}&=0,\label{eq:equil-u}\\
    \dfrac{\partial^2 W_{(1)}}{\partial x^2}+\dfrac{\partial^2 W_{(1)}}{\partial z^2}+\dfrac{\mathrm{d}\Psii}{\mathrm{d} x}-\dfrac{\partial p_{(1)}}{\partial z}&=0. \label{eq:equil-z}
\end{align}
\end{subequations}
By definition of the midline, ${U_{(1)}(x,0)=W_{(1)}(x,0)=0}$. The force-free boundary conditions~\eqref{eq:CauchyBC} read, at this order of the expansion,\begin{subequations}\label{eq:BC-surfaces}
\begin{align}
\left\llbracket\Psii + \frac{\partial U_{(1)}}{\partial z} + \frac{\partial W_{(1)}}{\partial x}\right\rrbracket_{z=\pm{h}/{2}}&=0, \\
\left\llbracket2\frac{\partial W_{(1)}}{\partial z}-p_{(1)}\right\rrbracket_{z=\pm {h}/{2}}&=0.
\end{align}
\end{subequations}
This provides six boundary conditions for Eqs.~\eqref{eq:ic1} and~\eqref{eq:equil}. Here, we impose the deformation of the midline $z=0$. This results in a force on the midline, so the solution need not be smooth across $z=0$. This allows us to impose all of these boundary conditions. This is not possible without this discontinuity for a thick plate, since the boundary conditions at $z=\pm h/2$ do not imply each other because the Kirchhoff hypothesis breaks down. An equivalent point has previously been made in Ref.~\cite{Piltner1992}.

\begin{figure*}[t]
\centering
\includegraphics[width=\linewidth]{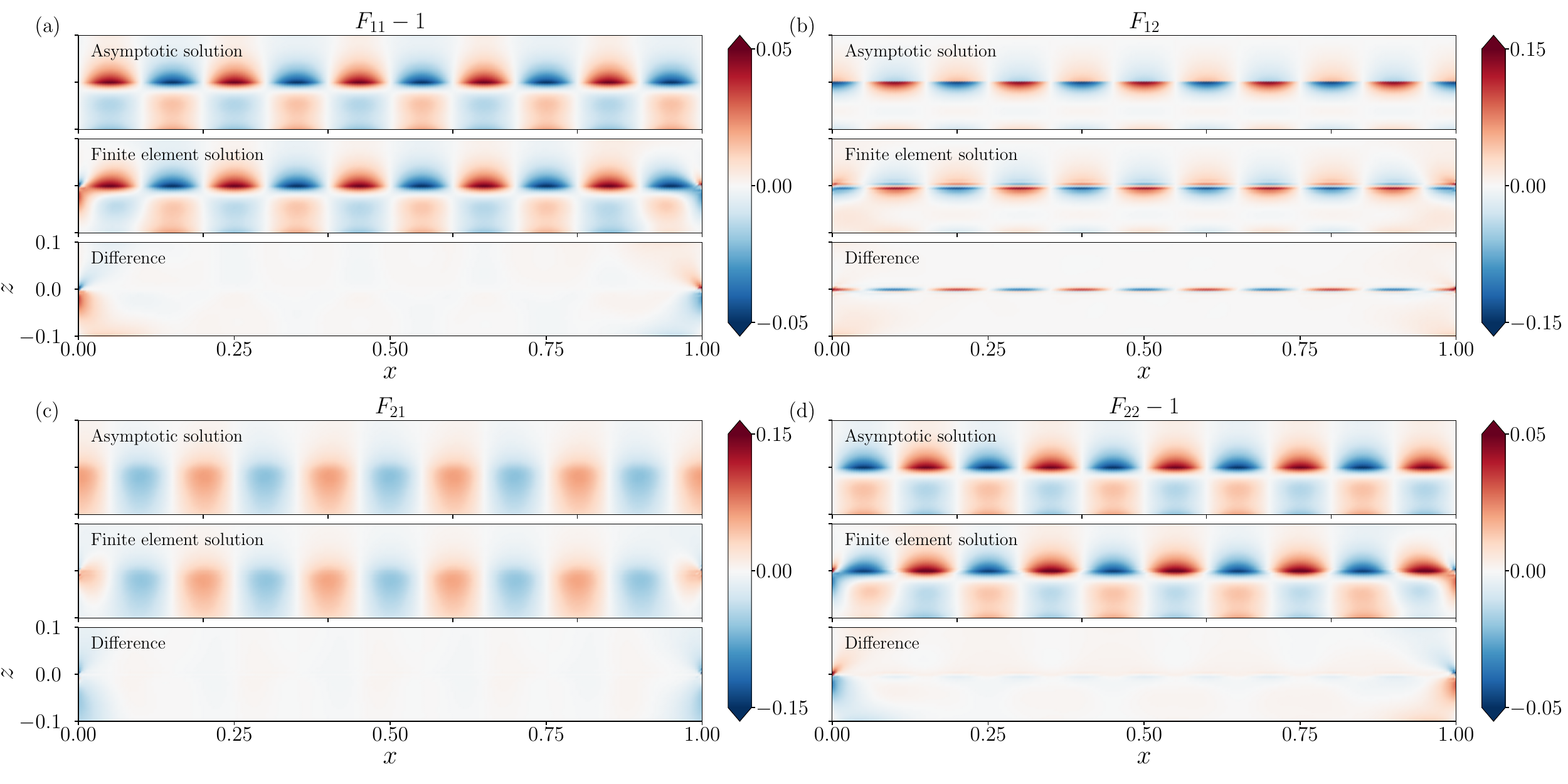}
\caption{Comparison between the asymptotic solution and finite-element simulations in \texttt{FEniCSx}~\cite{fenicsx} for a thick plate with imposed midline displacement. Plot of the components (a)~$F_{11}-1$, (b)~$F_{12}$, (c)~$F_{21}$, and (d)~$F_{22}-1$ of the deformation gradient in a representative Fourier mode: asymptotic solution (top), finite-element solution (middle), and their difference (bottom). Parameter values: $\varepsilon=0.1$, $h=0.2$, $k=10\pi$, $E_k=\Psii_k=0.5$.}
\label{fig2}
\end{figure*}

We solve this linear leading-order problem by considering a generic Fourier mode with wavenumber $k$, writing\begin{subequations}\label{eq:fourier expand}
\begin{align}
&E(x)= E_k\e^{\i k x}, &&\Psii(x) = \Psii_k \e^{\i k x},
\end{align}
and
\begin{align}
U_{(1)}(x,z) &= U_k(z) \e^{\i k x}, &W_{(1)}(x,z) &= W_k(z)\e^{\i k x}, \\
p_{(1)}(x,z) &= P_k(z) \e^{\i k x}.
\end{align}
\end{subequations}
Equations~\eqref{eq:ic1} and~\eqref{eq:equil} become
\begin{subequations}\label{eq:ODE-system}
\begin{align}
\i k U_k(z) + W_k'(z) + E_k &= 0,\label{eq:ODE-constraint}\\
U_k''(z) + \i k\bigl[E_k - P_k(z) + \i k U_k(z)\bigr]&= 0,\label{eq:ODE-U}\\
W_k''(z) - k^2 W_k(z) - P_k'(z) + \i k \Psii_k &= 0,\label{eq:ODE-W}
\end{align}
\end{subequations}
where dashes still denote differentiation with respect to the argument, now $z$. From Eqs.~\eqref{eq:ODE-constraint} and~\eqref{eq:ODE-U},
\begin{align}\label{eq:UPsol}
U_k(z) &= \frac{\i}{k}\bigl[E_k + W_k'(z)\bigr],&P_k(z) &= \dfrac{W_k'''(z)}{k^2}-W_k'(z).
\end{align}    
Substituting into Eq.~\eqref{eq:ODE-W} yields the fourth-order inhomogeneous differential equation
\begin{align}
W_k''''(z) - 2k^2W_k''(z) + k^4\,W_k(z) = \i k^3\,\Psii_k.
\label{eq:Wn-ODE}
\end{align}
for $z\in[-h/2,0]$ and $z\in [0,h/2]$. Finally, inserting Eqs.~\eqref{eq:UPsol} into Eqs.~\eqref{eq:BC-surfaces} yields the boundary conditions\begin{subequations}
\begin{align}
&W_k(0)=0,&&W_k'(0)=-E_k,
\end{align}
and
\begin{align}
&3k^2W_k'\left(\pm\frac{h}{2}\right) - W_k'''\left(\pm\frac{h}{2}\right) = 0, \\
&\i k^2 W_k\left(\pm\frac{h}{2}\right) + \i W_k''\left(\pm\frac{h}{2}\right) + k\Psii_k = 0.
\end{align}
\end{subequations}
The solution is~\cite{Note2}
\begin{align}
W_k(z)&=\dfrac{\i\Psii_k}{k}(1\!-\!\cosh{kz})-\dfrac{E_k}{k}\sinh{kz}+C_1^\pm kz\sinh{kz}\nonumber\\
&\quad+C_2^\pm(kz\cosh{kz}\!-\!\sinh{kz})\quad\text{for }z\gtrless 0,
\end{align}
where
\begin{subequations}
\begin{align}
C_1^\pm &= \frac{2\left[ \i\Psii_k (1 + \cosh{kh}) \pm E_k (kh + \sinh{kh}) \right]}{k\bigl(2 + k^2h^2 + 2\cosh{kh}\bigr)},\\
C_2^\pm &= \pm\frac{2\left[\i\Psii_k (kh - \sinh{kh})\mp E_k (1 + \cosh{kh}) \right]}{k\bigl(2 + k^2h^2 + 2\cosh{kh}\bigr)}.
\end{align}
\end{subequations}
Substituting this result into Eqs.~\eqref{eq:UPsol} determines the leading-order solution completely. We confirm this asymptotic solution by comparing it to finite-element simulations of a thick plate in \texttt{FEniCSx}~\cite{fenicsx}. This highlights the discontinuity across $z=0$~\wholefigref{fig2}. There is an excellent agreement between the asymptotic and finite-element solutions, with very small boundary effects~\footnote{Boundary effects arise from the additional boundary conditions on the lateral sides $x=0$, $x=1$ of the plate that we need in our finite-element simulations. We choose the force-free conditions $\tens{P}(0,z)\cdot(-\vec{e_x})=\tens{P}(1,z)\cdot\vec{e_x}=\vec{0}$.}.

\begin{figure}[b]
\centering
\includegraphics[width=\linewidth]{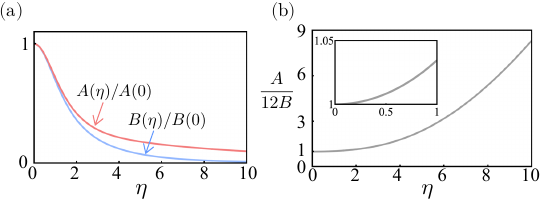}
\caption{Stretching and bending of (really) thick plates. (a)~Plot of the scaled effective stretching and bending moduli $A(\eta)/A(0)$ and $B(\eta)/B(0)$ against $\eta=kh$. (b)~Plot of the stretching number $\mathcal{S}=A/12B$ against $\eta$. Inset: zoom of this plot, showing that $\mathcal{S}$ changes by less than 5\% for $\eta\lesssim 1$.}
\label{fig3}
\end{figure}

\paragraph*{Thick plate theory.} Using our asymptotic solution, the elastic energy density~\eqref{eq:energy} has the formal expansion
\begin{align}
e=\dfrac{\varepsilon^2}{2}\sum_{k}{\sum_{\ell}{e_{k,\ell}(z)\e^{\i(k+\ell)x}}}+O\bigl(\varepsilon^3\bigr),
\end{align}
for some $e_{k,\ell}(z)$. The total elastic energy is therefore
\begin{subequations}
\begin{align}
\mathcal{E}=\dfrac{\varepsilon^2}{2}\int_0^1{\left(\sum_{k}{\sum_{\ell}{\mathcal{E}_{k,\ell}\e^{\i(k+\ell)x}}}\right)\mathrm{d}x}+O\bigl(\varepsilon^3\bigr),\label{eq:Eleading0}
\end{align}
in which
\begin{align}
\mathcal{E}_{k,\ell}=\int_{-h/2}^{h/2}{e_{k,\ell}(z)\,\mathrm{d}z}.\label{eq:Eklint}
\end{align}
\end{subequations}
Explicitly, from Eq.~\eqref{eq:energy},
\begin{align}
e_{k,\ell}(z)&=[\Psii_k+U_k'(z)+\i k W_k][\Psii_\ell+U_\ell'(z)+\i \ell W_\ell]\nonumber\\
&\qquad+4[E_k+\i k U_k(z)][E_\ell+\i \ell U_\ell(z)].\label{eq:ekl}
\end{align}
For $x\in[0,1]$, the wavenumbers $k,\ell$ take values in the set $\{2\pi n\mid n\in\mathbb{Z}\}$. Hence all terms in Eq.~\eqref{eq:Eleading0} with $k+\ell\neq 0$ vanish on integration over $x$. Thus
\begin{align}
\mathcal{E}=\dfrac{\varepsilon^2}{2}\sum_{\substack{k=2\pi n\\n\in\mathbb{Z}}}{\mathcal{E}_{k,-k}}+O\bigl(\varepsilon^3\bigr).\label{eq:Eleading}
\end{align}
We write $\tilde{\kappa}(x)=\varepsilon K(x)+O\bigl(\varepsilon^2\bigr)$, where $K(x)$ represents the leading-order bending strain. By definition, its Fourier amplitudes are ${K_k=\i k\Psii_k}$. With this, taking $\ell=-k$ in Eq.~\eqref{eq:ekl} and substituting into Eq.~\eqref{eq:Eklint}, we compute, using \textsc{Mathematica}~\cite{Note2},
\begin{align}
\mathcal{E}_{k,-k}=hA(kh)E_kE_{-k}+h^3B(kh)K_kK_{-k},\label{eq:Emodes}
\end{align}
where
\begin{subequations}
\begin{align}
&A(\eta) =\frac{4(\sinh{\eta}+\eta)}{\eta(2 + \eta^2 + 2\cosh{\eta})},\\
&B(\eta) = \frac{4(\sinh{\eta}-\eta)}{\eta^3(2 + \eta^2 + 2\cosh{\eta})}.
\end{align}
\end{subequations}
\paragraph*{Stretching and bending of really thick plates.} Equation~\eqref{eq:Emodes} recalls the structure of Eq.~\eqref{eq:plateE}: each of the modes contributing to Eq.~\eqref{eq:Eleading} separates into a stretching contribution and a bending contribution, proportional to $h$ and $h^3$, respectively. The prefactors $A$ and $B$ are thus effective stretching and bending moduli of the midline of the plate. (We note that, as for thin plates, there is no coupling between stretching and bending: terms proportional to $E_kK_{-k}, E_{-k}K_k$ vanish upon integration over $z$.)\looseness=-1

Both of these effective moduli depend on $\eta=kh$, i.e., on the ratio of the thickness $h$ of the plate to the length scale $k^{-1}$ over which the imposed midline deformation varies. The limit $\eta\to 0$, in which $A\to 2$, $B\to 1/6$, therefore corresponds to a thin plate; this recovers the asymptotic results for thin plates \cite{Haas21}. As $\eta$ increases, both $A,B$ decrease \figref{fig3}{a}, signalling softening of the plate to both stretching and bending.

The relative importance of stretching and bending is quantified by the stretching number $\mathcal{S}=A/12B$, so that $\mathcal{S}\to 1$ as $\eta\to 0$. As $\eta$ increases, so does $\mathcal{S}$ \figref{fig3}{b}. This implies that, unsurprisingly, for a thick plate, the energetic cost of stretching swamps that of bending. Nevertheless, $\mathcal{S}$ changes very little for $\eta\lesssim 1$ \figrefi{fig3}{b}, which results from the small prefactor in the leading-order expansion
\begin{align}
\mathcal{S}-1=\dfrac{\eta^2}{30}+O\bigl(\eta^4\bigr).
\end{align}
Not only does the separation of stretching and bending thus arise for small deformations of a really thick plate, too, but also the relative importance of stretching and bending remains (essentially) unchanged from the case of a thin plate, up to deformations over lengthscales comparable to the thickness of the plate. This provides an explanation for the surprising accuracy of thin-plate theories that has motivated our calculation.

In our work, we have studied a new asymptotic limit for elastic plates: rather than describing finite deformations of asymptotically thin plates, we have imposed asymptotically small deformations of the midline a really thick elastic plate. We have discovered that the asymptotic separation of stretching and bending that is the hallmark of thin-plate theories also arises in this asymptotic limit. The associated effective stretching and bending moduli soften with increasing thickness. Still, even for deformations over lengthscales comparable to the thickness of plate, the relative contributions of stretching and bending are almost the same for thin plates and for our thick plates. Thin-plate theories can therefore capture the mechanical behaviour of these thick plates, up to rescaling external forces to absorb the softening moduli.

Derivations of thin-plate theories continue to attract attention~\cite{Wang2016,Karttunen2017,Wood2019,Ozenda2021,Vitral2022}. In particular, there has been a lot of recent interest in theories of morphoelastic plates and shells~\cite{Dervaux2009,Mcmahon2011,Moulton2013,sadik16,Moulton2020,Haas21,Andrini2025,Taffetani2026} that add (biological) growth to the classical theories, related models of non-Euclidean plates~\cite{Efrati2009,Pezzulla2017}, and even curved metamaterials~\cite{Giomi26}. In morphoelasticity, the separation of stretching and bending gives rise to notions of intrinsic stretching and intrinsic bending that can represent, biologically, cell shape changes during development. This idea has been applied in particular to the developmental mechanics of the green alga \emph{Volvox}~\cite{Hohn2015,Haas2018a,Haas2018b,Chen2024,Haas2025}. How the additional nonlinear effects, including coupling of stretching and bending~\cite{Haas21}, uncovered by some of these extended thin shell theories translate to thick shells remains an open question. Still, the cell sheet of \emph{Volvox} and of related algae in the family Volvocaceae~\cite{Tribet2026} is not thin. Our asymptotic results therefore provide a justification for deploying the picture of stretching and bending energies on descriptions of the mechanics of this biological system.

We have assumed, here, the simplest, neo-Hookean constitutive relations, expressed by Eq.~\eqref{eq:E}. They are the leading-order term in a polynomial expansion of a general, two-dimensional, isotropic, incompressible hyperelastic energy density $W(\mathcal{I})$ in powers of $\mathcal{I}-2$, where $\mathcal{I}=\tr{\tens{F}^\top\tens{F}}$. Our calculations shows that $\mathcal{I}=2+O\bigl(\varepsilon^2)$, so higher-order in this expansion do not contribute to the leading-order plate theory. This assumption does not therefore lose generality. The same is true in leading-order theories of thin plates~\cite{Dervaux2009,Haas21}: a general \emph{three-dimensional}, isotropic, incompressible  hyperelastic energy density has a polynomial expansion in powers of $\mathcal{I}_1-3$, $\mathcal{I}_2-3$, where $\mathcal{I}_1=\tr{\tens{C}}$ and ${\mathcal{I}_2=\bigl(\mathcal{I}_1^2-\tr{\tens{C}^2}\bigr)/2}$, with $\tens{C}=\smash{\tens{F}^\top\tens{F}}$, are the principal invariants \cite{goriely}. Terms beyond the neo-Hookean term do not contribute at the order $O\bigl(\varepsilon^2\bigr)$ of the leading-order thin-plate theory because $\mathcal{I}_1=\mathcal{I}_2 +\smash{O\bigl(\varepsilon^3\bigr)}$ \cite{Haas21}. In this context, motivated by the extreme softness of biological materials~\cite{sheiko19,lou23}, we have recently introduced ``purely nonlinear'' elasticity at zero (linear) shear modulus~\cite{zhao25}. The mechanical response of these materials is dominated by such higher-order terms. The corresponding thin-shell theory involves stretching and bending energies proportional to $E^4$ and $K^4$, respectively~\cite{zhao25}, but the corresponding thick shell picture remains another open problem.

Finally, and more generally, our work opens up one more question: are there other asymptotic regimes, beyond the thin-plate limit and the limit of small deformations of a thick plate that we have introduced here, in which an asymptotic derivation of a plate theory is possible and so new insights into the behaviour of elastic plates and shells can emerge?\\

\begin{acknowledgments}
We gratefully acknowledge funding from the Max Planck Society. 
\end{acknowledgments}

\subsection*{{Data availability}}
A \textsc{Mathematica} notebook for the asymptotic calculations is provided in the Supplemental Material~\cite{Note2}. \texttt{FEniCSx}~\cite{fenicsx} code for the finite-element simulations is available openly at Ref.~\footnote{Code is available at \href{https://doi.org/10.5281/zenodo.20413363}{\texttt{doi.org/10.5281/zenodo.20413363}}.}.

\bibliography{main.bib}
\end{document}